\documentstyle[12pt,epsfig]{article}

\newcommand{\be}{\begin{equation}}
\newcommand{\ee}{\end{equation}}
\newcommand{\ba}{\begin{array}}
\newcommand{\ea}{\end{array}}
\newcommand{\baa}{\begin{eqnarray*}}
\newcommand{\btab}{\begin{tabular}}
\newcommand{\etab}{\end{tabular}}
\newcommand{\eaa}{\end{eqnarray*}}

\begin{document}

\begin{titlepage}
\begin{flushright}
\begin{tabular}{l}
CERN--TH/98--290\\
NORDITA--98--58--HE\\
hep-ph/9809287
\end{tabular}
\end{flushright}

\vskip2cm
\begin{center}
  {\large \bf
             The Operator Product Expansion,\\
             Non-perturbative Couplings and the Landau Pole:\\
             Lessons from the $O(N)$ $\sigma$-Model 
  \\}

\vspace{2cm}
{\sc M.~Beneke}${}^1$, {\sc V.M.~Braun}${}^{2,3}$
          and {\sc N.~Kivel}${}^3$
\\[0.8cm]
\vspace*{0.1cm} ${}^1$ {\it
          Theory Division, CERN, CH-1211 Geneva 23, Switzerland 
                       } \\[0.2cm]
\vspace*{0.1cm} ${}^2${\it 
          NORDITA, Blegdamsvej 17, DK-2100 Copenhagen, Denmark
                       } \\[0.2cm]
\vspace*{0.1cm} ${}^3$ {\it 
          St. Petersburg Nuclear Physics Institute,
          188350 Gatchina, Russia
                       } \\[1.0cm]

\vskip2.1cm
{\bf Abstract:\\[10pt]} \parbox[t]{\textwidth}{ 
We obtain the operator product expansion of the self-energy 
in the $O(N)$ non-linear $\sigma$-model to all orders in the 
coupling and the large momentum, and to next-to-leading order in $1/N$. 
In the light of this result we discuss recent suggestions that 
there may be additional power corrections from short distances, 
associated with defining the coupling constant 
non-perturbatively. The non-linear $\sigma$-model provides no 
evidence for such `non-standard' power corrections. We also find that 
the OPE converges for sufficiently large external momentum, 
presumably because there are no multi-particle thresholds at
arbitrarily high energies in the $1/N$ expansion.
}
\vskip1cm 

\end{center}

\end{titlepage}

\noindent
{\bf 1.}  Short-distance observables in QCD, 
characterized by a large scale $Q\gg \Lambda_{\rm QCD}$, can 
be predicted perturbatively in $\alpha_s(Q)\ll 1$. In some cases 
non-perturbative power corrections in $\Lambda_{\rm QCD}/Q$ 
can be incorporated as well, using the operator product expansion (OPE) 
\cite{Wil69,SVZ79}. Alternatively, one can deduce the structure of 
non-perturbative effects by investigating the infrared (IR) contributions
to loop integrals in perturbation theory. This the basis
of the so-called renormalon method \cite{BB}.
Either way, for example 
for current two-point functions $i\int\!dx \,e^{i qx} \langle 0|
T(j(x) j(0))|0\rangle$, where $j$ is a bilinear quark current, 
one finds that the leading power correction scales as 
$(\Lambda_{\rm QCD}/Q)^4$ and is related to the gluon condensate. 
Both methods parametrize power corrections that arise from long-distance 
contributions to the observable. In general one cannot exclude 
non-perturbative effects from short distances. The only known dynamical 
mechanism for such corrections is due to instantons of 
small size $\rho^{-1}\gg\Lambda_{\rm QCD}$ \cite{inst}. They scale as 
$(\Lambda_{\rm QCD}/Q)^{11-2N_f/3}$ and are strongly suppressed 
at large $Q$.

In recent work Grunberg \cite{Gru97} and Akhoury and Zakharov \cite{AZ97a}  
suggested that one should expect quite 
generically power corrections of order 
$(\Lambda_{\rm QCD}/Q)^2$ from short distances on grounds that the 
perturbative QCD coupling is usually not specified to an accuracy better 
than this. To make their point let us consider the integral
\begin{equation}
\label{integral}
\int dk^2\,F(k,Q)\,\frac{\alpha_s(k)}{k^2}
\label{quest}
\end{equation}
which is typical of renormalon calculations of diagrams with one gluon 
line. When the loop momentum $k$ is small, and if $F(k,Q)$ behaves as 
$(k/Q)^p$ for small $k$, one obtains a contribution of order 
$(\Lambda/Q)^p$ to the integral. This term stands for the power correction 
that is taken into account by the OPE or the renormalon method. In this 
context the definition of the coupling $\alpha_s$ has played no role. 
It is usually considered to be the one-loop perturbative running coupling, 
which has a Landau pole at $k^2=\Lambda_{\rm QCD}^2$. However, physical 
quantities do not have a Landau pole and a non-perturbative definition 
of $\alpha_s(k)$, valid at all $k$, should not have a Landau pole either. 
Such a non-perturbative coupling can itself be expanded in a power 
series, when $k$ is large:
\begin{equation}
\label{coupling}
\alpha_s(k) = \frac{1}{-\beta_0\ln(k^2/\Lambda_{\rm QCD}^2)} + 
\ldots + \,\mbox{const}\cdot \frac{\Lambda_{\rm QCD}^2}{k^2} +
\ldots.
\end{equation}
The dots denote further perturbative and power correction terms. If the 
second term is present, and if one insists on extending  
(\ref{quest}), derived in the context of perturbation theory,  
to a non-perturbative 
running coupling, one obtains a power correction of order 
$\Lambda_{\rm QCD}^2/Q^2$ to the integral in (\ref{integral}) 
from the region of large $k\sim Q$. 
In \cite{Gru97,AZ97a} these additional power corrections are interpreted 
as a potential new source of non-perturbative effects, not accounted 
for by the OPE, but also not in contradiction with the OPE, 
because they arise from short distances (large $k$). 
The concept of an effective 
coupling is not unique and the existence of a particular $1/k^2$ term 
in (\ref{coupling}) is not compelling. 
Independent of the precise form of power corrections, however, 
the argument given in \cite{Gru97,AZ97a} suggests the existence of 
unconventional power corrections at some order.

The present work is motivated by our attempt to understand this  
argument, and its relation to the discussion of a related issue 
in \cite{BBB95}, in the context of  
the  two-dimensional non-linear $O(N)$ $\sigma$-model. 
We calculate the asymptotic expansion for the self-energy of the 
$\sigma$ particle to next-to-leading order in $1/N$ and to all powers in the 
large momentum. 
The approximation contains 
all non-trivial elements of the OPE construction in QCD, except for 
the absence of multi-particle production. In particular, it allows us 
to address the question of non-perturbative corrections to the 
effective coupling and potential power corrections from short distances. 
Our conclusions can be summarized as follows: 

(a) There are power 
corrections from the region $k\sim Q$, but they are included in the standard 
OPE construction and correspond to soft subgraphs of the graphs that 
define the effective coupling non-perturbatively. Despite the fact that 
$k\sim Q$, these power corrections are therefore of infrared origin. 

(b) The fact that the perturbative coupling has a 
Landau pole does not give rise to additional power corrections. The 
Landau pole disappears and the correct analyticity structure is recovered 
only after summation of the OPE. The issue of defining an 
effective coupling non-perturbatively is 
not related to the existence of power corrections to 
particular physical observables. In fact, it 
turns out that the OPE is naturally expressed in terms of the 
perturbative coupling (the first term on the right-hand side of 
(\ref{coupling})). 

(c) The OPE for the self-energy is convergent for 
$|Q^2| > 9 m^2$ in the complex $Q^2$ plane, where $m$ is the dynamically 
generated mass of the $\sigma$ particle, analogous to $\Lambda_{\rm QCD}$ 
in QCD. The radius of convergence is determined by the onset of 
the 3-particle cut. The convergence of the OPE 
is probably a consequence of the $1/N$ expansion, because to 
the order we are working at most three particles can be 
produced. More generally, 
we expect that the OPE of the self-energy is convergent 
for $|Q^2|>(2k+1)^2 m^2$ if 
the $1/N$ expansion is truncated at the order $k$. 
Thus, for finite $N$, we expect the OPE to be divergent for any 
$Q^2$. Because of the restriction on the number 
of produced particles, the $1/N$ expansion in the $\sigma$-model 
is not suited for studies of the analytic continuation of the OPE 
to Minkowskian momenta and the validity of ``parton-hadron duality'', 
and we do not address both questions in this work.

We emphasize that while the present calculation clarifies the 
argument of \cite{Gru97, AZ97a} it does not preclude the existence of 
power corrections from short-distances due to a yet unknown 
dynamical mechanism other than small instantons, see \cite{AZ97b}
for a particular suggestion. Up to this date, however, all (known) 
short-distance power corrections are of semiclassical origin  
and unrelated to integrals of the type (\ref{quest}) and the 
issue of defining the coupling constant.\\

\noindent
{\bf 2.} The nonlinear $O(N)$ $\sigma$-model was considered repeatedly
as a toy model for the OPE \cite{david,NSVZ,terentev}. 
In particular, the non-perturbative separation of long-distance and 
short-distance contributions is well understood in this model, both 
in cut-off schemes \cite{NSVZ} and in dimensional regularization 
\cite{david}. In the second factorization scheme, the coefficient 
functions have non-Borel-summable perturbative expansions due to infrared 
(IR) renormalons. The prescription dependence that follows from defining 
the series is exactly compensated by the prescription dependence 
of defining the power-divergent matrix elements of higher-dimension 
operators. In the subsequent calculation we observe this compensation 
to all orders in the OPE. However, in this Letter we are mainly concerned 
with power corrections not related to IR renormalons, in particular 
the effects of a coupling redefinition on the power series expansion, 
and with analytic properties of the power series truncated to finite order.

The construction of the $1/N$ expansion and the Feynman rules are detailed 
in \cite{david, NSVZ}. The $N$ $\sigma$-particles have 
mass\footnote{To avoid factors of $4\pi$, we rescale the coupling by 
a factor $4 \pi$. With this convention $g$ satisfies 
\begin{displaymath}
\beta(g)=\mu^2\frac{\partial g}{\partial \mu^2} = -g^2.
\end{displaymath}
Note that this $\beta$-function is exact in leading order of the 
$1/N$ expansion. In particular, $g(k)$ has a Landau pole at 
$k^2=m^2$.}
\begin{equation}
m^2=\mu^2\,e^{-1/g(\mu)},
\end{equation}
where $\mu$ is a renormalization scale. 
In addition to the standard massive propagator for the scalar 
$\sigma$-particles, the $1/N$ expansion involves the propagator of the 
auxiliary ``$\alpha$-particle'' 
\begin{equation}
D(k) = -4\pi \,(k^2+2m^2)\,g_{\rm eff}(k) = 
-4\pi\sqrt{k^2(k^2+4m^2)}\,\, \ln^{-1}
\left[\frac{\sqrt{k^2+4m^2}+\sqrt{k^2}}
{\sqrt{k^2+4m^2}-\sqrt{k^2}}\right].
\label{D}
\end{equation}
In order to establish a connection with (\ref{quest}), 
we interpret $g_{\rm eff}(k)$ defined through the 
$\alpha$-propagator as above as a non-perturbative effective coupling.
The factor $k^2+2m^2$ is separated such that
$g_{\rm eff}(k)$ can be viewed as the sum of the chain of $\sigma$-particle 
bubble graphs with momentum-dependent vertices
$$
g_{eff}(k) = \frac{1}{1/g_0-I(k,m)}=\\[4mm] g_0+\
g_0^2
\mskip-5mu
\ba{l}
\\[-3mm]
\epsfig{file=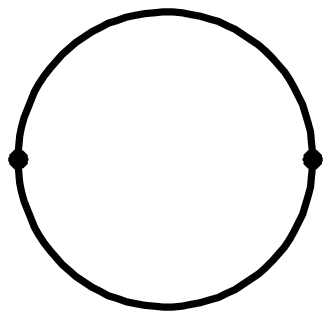,height=1.5cm,width=1.5cm,clip=}
\ea
\mskip-5mu
+\
g_0^3\,\,
\mskip-5mu
\ba{l}
\\[-3mm]
\epsfig{file=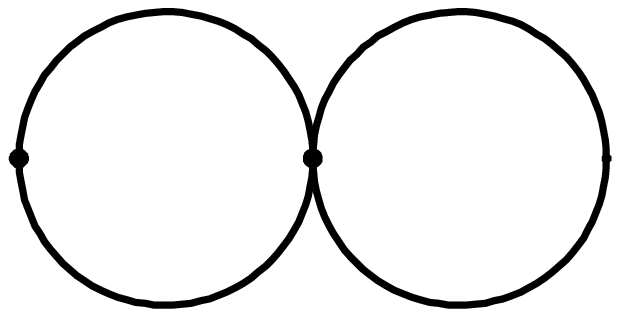,height=1.5cm,width=2.2cm,clip=}
\ea
+ \ldots,
$$
\\[-5mm]
where an UV regularization is implied, $g_0$ is the bare coupling and
\begin{equation}
I(k,m)=\int d^2q \frac{q(q+k)}{[q^2+m^2][(q+k)^2+m^2]}.
\end{equation}
The coupling $g_{\rm eff}(k)$ can be expanded at large momenta
generating a power series
\begin{equation}
\label{geff}
g_{\rm eff}(k) = g(k) -2g^2(k) \frac{m^2}{k^2} + \ldots.
\end{equation}
Following the terminology of \cite{Gru97}, $g_{\rm eff}(k)$ 
naturally splits into a perturbative and 
non-perturbative part, $g_{\rm eff}(k)= g(k)+\delta g(k)$, where 
the perturbative part $g(k)$ is given by the first term on 
the right-hand side alone. 

It is important to note that the power corrections $\delta g(k)$ 
to the running coupling are of IR origin. Diagrammatically, 
the reason for this is that even if the momentum that flows into the chain 
of bubbles is large, one of the lines in any one or more of the 
bubbles can still be soft.
 
$D(k)$ does not have a Landau pole, but a cut at $-k^2>4 m^2$. 
Because $k^2+2m^2$ is factored out, $g_{\rm eff}(k)$ has 
a `kinematic' singularity at $-k^2 = 2 m^2$ in addition. Note, however, 
that whenever the power expansion of the effective coupling is truncated, 
there is an additional Landau singularity at $k^2=m^2$.\\

\begin{figure}[t]
   \vspace{-4.8cm}
   \epsfysize=26cm
   \epsfxsize=18cm
   \centerline{\epsffile{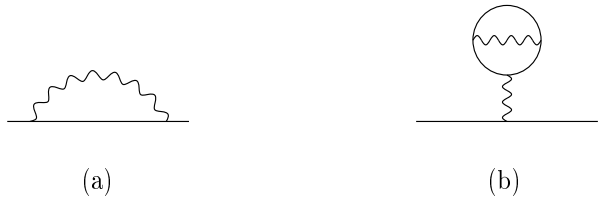}}
   \vspace*{-18cm}
\caption[dummy]{\small $\sigma$-self-energy diagrams at 
order $1/N$. \label{fig1}}
\end{figure}

\noindent
{\bf 3.} Consider the propagator of a $\sigma$-particle with 
momentum $p^2\gg m^2$. To leading order in $1/N$ it is equal to 
$(p^2+m^2)^{-1}$ and we can interpret its expansion at large momenta
as the operator product expansion 
\begin{equation}
 (p^2+m^2)^{-1} = 
\frac{1}{p^2}\sum_{n=0}^\infty \left(\frac{-m^2}{p^2}\right)^n \equiv
\frac{1}{p^2}\sum_{n=0}^\infty C^{(0)}_n(p^2)\langle O_n\rangle^{(0)},
\end{equation} 
with $C^{(0)}_n(p^2)=-1/p^{2n}$ and $\langle O_n\rangle^{(0)} = m^{2n}$.
The OPE interpretation implies that the terms
identified as operator matrix elements originate from large distances.  
To see this, one has to start from the perturbative phase of the $\sigma$ 
model and construct the OPE explicitly; details can be found in \cite{NSVZ}. 
We shall take the OPE interpretation of the large-$N$ result for granted, 
and concentrate on the first subleading order in the large $N$-expansion,
formulated in terms of massive $\sigma$-fields already.

To $1/N$ accuracy the $\sigma$-propagator involves the 
self-energy correction given by the two diagrams shown in 
Fig.~\ref{fig1}, where wavy lines denote the $\alpha$-propagator (\ref{D}). 
The self-energy is quadratically ultraviolet divergent. 
We define the renormalized self-energy by zero-momentum subtractions
\begin{equation}
\Sigma^{\rm ren}(p) =  \Sigma(p) -\Sigma(0) 
-p^2\frac{\partial}{\partial p^2}\,\Sigma(p)_{\big|p^2=0} 
\label{Sren}
\end{equation}
and drop the superscript ``ren'' in what follows. The tadpole diagram 
in Fig.~\ref{fig1}b is then subtracted completely, and the self-energy 
is given by 
\begin{equation}
\label{sig}
\Sigma(p) = \frac{1}{\pi N}\int \!d^2 k 
          \,\,\frac{k^2+2m^2}{(p+k)^2+m^2}\,\,g_{\rm eff}(k)\,
+\,\mbox{\rm subtractions}.
\end{equation}
We have to verify that the asymptotic expansion of the propagator
can be written in the factorized form  
\begin{eqnarray}
  \lefteqn{(p^2+m^2)^{-1} - (p^2+m^2)^{-2} \Sigma(p) =}
\nonumber\\ &=& 
 \frac{1}{p^2}\sum_{n=0}^\infty \Big[C^{(0)}_n(p^2)+(1/N)C^{(1)}_n(p^2)\Big]
\Big[\langle O_n\rangle^{(0)} +(1/N)\langle O_n\rangle^{(1)}\Big]+ O(1/N^2)   
\end{eqnarray} 
where $C^{(1)}_n(p^2)$ and $\langle O_n\rangle^{(1)}$
correspond to $1/N$ corrections to coefficient 
functions and matrix elements, respectively. Note that the explicit 
construction of the operators is not necessary and is 
complicated in the case at hand by the existence of infinitely 
many operators of the same dimension. The OPE statement means that 
the contributions from small and large distances to (\ref{sig}) can be 
factorized. In the present case this also means that all contributions 
from large distances are associated with operator matrix elements.

Returning to the integral (\ref{sig}), long-distance contributions 
originate from the IR sensitive regions $k^2\sim m^2$ and 
$(p+k)^2\sim m^2$. The OPE is obtained by expanding the integrand of 
(\ref{sig}) in small quantities given a particular loop momentum 
region. The following three loop momentum regions have to be 
considered separately:
\begin{itemize}
\item[(i)] $k\gg m$, $p+k\gg m$: Both lines are far off-shell and both 
the $\sigma$-propagator and the effective coupling can be expanded in 
$m^2$. The diagram effectively contracts to a point and the result can 
be interpreted as a $1/N$ correction to the short-distance coefficients 
times operator matrix elements evaluated at leading order in $1/N$.
\item[(ii)] $k\sim m$, $p+k\gg m$: The $\sigma$-propagator can be expanded, 
but the effective coupling cannot. Thus the $\sigma$ propagator
is contracted, but the $\alpha$-propagator is not; the resulting diagram 
can be viewed as a $1/N$-correction to the vacuum expectation value
of the product of two $\alpha$-fields $\langle0|\alpha(x)\alpha(0)|0\rangle$
with small separation $x\sim 1/p$. Expansion in $x$ yields as the leading 
contribution a $1/N$-correction to the matrix element 
$\langle0|\alpha^2|0\rangle$ times a leading-order short-distance 
coefficient. (This matrix element was evaluated explicitly in 
\cite{david,NSVZ}.) Note that from power counting it follows that this 
region contributes to (\ref{sig}) first at order $p^2\cdot m^4/p^4$, 
i.e. with a $1/p^4$-suppression similar to the gluon condensate in 
QCD. (The argument does not hold for the contribution of the 
subtraction terms.)
\item[(iii)] $k\gg m$, $p+k\sim  m$: The effective coupling can be expanded, 
but the $\sigma$-propagator cannot be expanded. This case looks symmetric 
to the previous one but the interpretation is different: contracting 
the $\alpha$-propagator one gets a {\em leading} 
contribution at large $N$ to the vacuum expectation
value for the product of two $\sigma$-fields 
$\langle0|\sigma^a(x)\sigma^a(0) |0\rangle =  1/g(x^{-1})\cdot 
[1+m^2x^2/4+\ldots]$ (no sum over `$a$'). Note the inverse 
coupling as a consequence of the normalization of the ``length'' of the 
$\sigma$-field. The result can therefore be interpreted as a $1/N$-correction 
to the short-distance coefficient times a leading large $N$ 
operator matrix elements of $\sigma$ fields. Note that a contribution 
of this type is usually not considered in the renormalon method. 
However, it could be accounted for as well. In the QCD analogue of 
our problem, this would correspond to renormalon 
contributions to the coefficient function of a quark condensate. 
\end{itemize}
We emphasize that all three contributions are part of the usual 
OPE construction and that (i,iii) include the ``short-distance'' 
power corrections discussed in \cite{Gru97,AZ97a}. As already mentioned, 
these power corrections are in fact also of long-distance origin.  

For the actual calculation of the expansion of (\ref{sig}) one could 
follow the above procedure and calculate the contribution from each region 
separately. It is more convenient not to perform this split-up and to 
calculate the expansion directly. To this end we write the 
inverse logarithm in $D(k)$ as 
\begin{equation}
\label{at}
\int_0^\infty\!d t \,\left[\frac{A-1}{A+1}\right]^{t},
\end{equation}
where $A=(1+4 m^2/k^2)^{1/2}$. Subsequently, we use the Mellin-Barnes 
representation
\begin{equation}
A\,\left[\frac{A-1}{A+1}\right]^{t} = \int\frac{ds}{2\pi i}\,
K(s,t)\left(\frac{m^2}{k^2}\right)^{-s}, 
\end{equation}
where the kernel $K(s,t)$ is given by 
\begin{equation}
K(s,t) = \frac{\Gamma(t+s)\,\Gamma(-1-2 s)}{\Gamma(t-1-s)}
\left[1+\frac{2 \,(t+s)}{t-s-1}+\frac{(t+s) (t+s+1)}{(t-s-1) (t-s)}
\right].
\end{equation}
After these manipulations all integrations can be done and we arrive 
at the following ``Borel representation'' for the expansion in 
powers (and logarithms) of $m^2/p^2$: 
\begin{equation}
\Sigma(p) = \frac{p^2}{N} \int_0^\infty\! dt\,
\sum_{n=0}^\infty \left(-\frac{m^2}{p^2}\right)^n
\Bigg\{e^{-t/g(p)}\Bigg[F_{\rm p}^{(n)}[t]\frac{1}{g(p)}
+G_{\rm p}^{(n)}[t]\Bigg] - H_{\rm np}^{(n)}[t]\Bigg\}. 
\label{SE}
\end{equation}
Explicit expressions for the functions  $F_{\rm p}^{(n)}[t]$, 
$G_{\rm p}^{(n)}[t]$, $H_{\rm np}^{(n)}[t]$ are given in the appendix.
All dependence on $m^2/p^2$ is either explicit or implicit in 
$g(p)=1/\ln(p^2/m^2)$. 

We first explain the origin of the various terms in (\ref{SE}).

The functions $F_{\rm p}^{(n)}[t]$ and $G_{\rm p}^{(n)}[t]$ correspond 
to the contributions of regions (i,iii) discussed above. Because they 
are multiplied by the factor $e^{-t/g(p)}$, they can be interpreted 
as Borel transforms of the perturbative expansions of short-distance 
coefficients times matrix elements that do not depend on $g(p)$. 
The separation of regions (i) and (iii) is not unique. In region (i) 
the expansion of the $\sigma$-propagator in $m^2/(p+k)^2$ leads to 
IR divergences, which require an intermediate regularization. The result 
from region (i) then takes the form
\begin{equation}
F_{\rm p}^{(n)}[t]\,\ln\frac{p^2}{\mu^2}+G_{\rm p}^{(n,i)}[t],
\end{equation}
where $\mu$ is the intermediate factorization scale. On the other hand, 
the expansion of $g_{eff}$ in region (iii) leads to increasingly UV 
divergent integrals, for which the same intermediate regularization 
is required. The result 
from region (iii) then takes the form
\begin{equation}
F_{\rm p}^{(n)}[t]\,\ln\frac{\mu^2}{m^2}+G_{\rm p}^{(n,iii)}[t].
\end{equation}
Both contributions combine to the square bracket in (\ref{SE}). The 
intermediate (and arbitrary) scale $\mu$ drops out and the logarithms 
combine to the inverse running coupling. Recall that region (iii) is 
a long-distance contribution to the integral (\ref{sig}), but it is 
multiplied by a perturbative expansion, because large momentum flows 
through the effective coupling. After combining (i) with (iii) as 
in (\ref{SE}), both $F_{\rm p}^{(n)}[t]/g(p)$ and $G_{\rm p}^{(n)}[t]$ 
have long- and short-distance contributions. 

The function $H_{\rm np}^{(n)}[t]$ 
originates from region (ii) discussed above and therefore should be 
identified with the $1/N$ correction to operator matrix elements. It 
cannot be written as a series expansion in $g(p)$ and has only 
long-distance contributions. No extra regularization is needed 
to separate (ii) from (i,iii). 

Nevertheless, the separation of (ii) from (i,iii) is not unique. 
The functions $F_{\rm p}^{(n)}[t]$, $G_{\rm p}^{(n)}[t]$, 
$H_{\rm np}^{(n)}[t]$  
have singularities (typically simple poles) in the complex $t$ plane 
at integer values $t=\pm k$, $k=1,2,\ldots$. These are just the 
usual ultraviolet and IR renormalon singularities. 
(The poles at $t=0$ correspond to the usual logarithmic 
UV divergences and are cancelled by renormalization.) 
Using the expressions collected in the appendix, one can verify that 
all IR renormalon singularities at positive values of $t$ cancel. 
The cancellation of a particular singularity at $t= t_0$ occurs 
between $G_{\rm p}^{(n)}[t]$  and $H_{\rm np}^{(n+t_0)}[t]$ and thus 
involves a cancellation between a short-distance coefficient 
and an operator matrix element over different orders in the power 
expansion. This is an explicit all-order verification of the cancellation 
mechanism discussed in \cite{david}. As a consequence of the singularities 
in individual terms of the sum over $n$, the summation and 
the integration over $t$ cannot be interchanged, unless the integration 
contour is shifted slightly above (or below) the real axis. 
Only after such a definition can one truncate the OPE. Obviously 
the prescription dependence cancels in the end.

As an explicit example, consider the leading asymptotic behaviour 
of $\Sigma(p)$ as $p\to\infty$, related to the expansion of 
$F_{\rm p}^{(n=0)}[t]$ and $G_{\rm p}^{(n=0)}[t]$ around $t\to 0$ and to 
$H_{\rm np}^{(n=0)}[t]$. We find
\begin{equation}
 \Sigma(p)= \frac{p^2}{N}\left[ \ln g(p)+ 1.887537 - 2\, g(p) +
       \sum_{n=1}^\infty \sigma_n\, g^{n+1}(p)\right],
\end{equation}
with factorially divergent coefficients
\begin{equation}
 \sigma_n = n!\left\{[1+(-1)^n]\zeta(n+1)-2\right\}.
\label{PT}  
\end{equation}
Because of the Riemann zeta-function this gives rise to an infinite 
series of renormalon poles. 

Returning to (\ref{SE}), we note that the OPE is naturally expressed 
in terms of the perturbative coupling $g(p)$.
The reason is that the asymptotic expansion is mathematically
an expansion in  powers and logarithms of $m^2/p^2$,  and 
absorbing logarithms only into the coupling leaves all powers 
of $m^2/p^2$ explicit. This implies that at every finite order in the 
OPE there is a Landau singularity at $p^2=m^2$, which disappears, 
when the expansion is summed. 

One can avoid unphysical Landau singularities 
order by order in the asymptotic expansion, 
if one eliminates  $g(p)$ in favour of 
$g_{eff}(p)$, which amounts to an all-order re-arrangement of the OPE. 
However, it is important to stress that there is no advantage to 
doing this: although the Landau singularity never appears, the 
predictive power is not enlarged. One would still have to sum the 
entire OPE to be able to predict the correlation function accurately 
in the region $p^2\sim m^2$. In general, 
(but not in the case of $g_{eff}(p) $), 
an inadequate choice of the non-perturbatively 
defined coupling may obscure the cancellation of renormalons poles, 
or even introduce power corrections into the OPE which are 
not natural for the physical quantity in question, but which 
appear in order 
to cancel power corrections implicit in the coupling definition
(see the discussion in \cite{BBB95}). 

In the present example all power corrections 
do indeed originate from long distances and can be accounted for in 
the OPE framework. The calculation clarifies that the fact that the 
perturbative coupling has a Landau pole, or that a non-perturbative 
coupling has itself power corrections does not provide sufficient 
motivation for introducing ad hoc power corrections beyond 
the OPE. Moreover, all known sources of short-distance power 
corrections are related to semiclassical effects.
Semi-classical effects are connected with 
multi-particle emission, a phenomenon which is 
certainly beyond the approximation 
in which one assumes the emission of one quantum gluon
through an effective coupling. It therefore seems that to address 
the question of whether the OPE is complete, one would have to 
abandon the effective coupling representation. 

A short comment is in order on  
ultraviolet renormalons, which lead to factorially divergent, 
sign-alternating series in the $\sigma$-model (as in QCD). This 
divergence limits the accuracy of a purely perturbative 
approach and the best possible approximation has an error that scales 
as a power in $m^2/p^2$. It is sometimes argued (see for example 
\cite{AZ97a}) that UV renormalons require adding power corrections 
(of short-distance nature) to the OPE. In our example, the series 
expansion of the leading term (\ref{PT}) as well as all others lead 
to UV renormalon singularities at negative $t$. Being located at negative 
$t$, these singularities do not lead to ambiguities in the individual terms 
of (\ref{SE}). Since the OPE (\ref{SE}) reproduces the exact result 
(as one can check numerically), we also conclude that UV renormalons 
do not necessitate introducing additional power corrections, as expected 
on general grounds for sign-alternating series.\\

\noindent
{\bf 4.} 
Given the OPE to all orders, it is interesting to ask about the 
convergence/divergence properties of the OPE itself (rather than the 
perturbative expansion that multiplies each power correction). For 
example, in \cite{Shi94} a toy heavy-quark correlation function is 
used to demonstrate that in this case the heavy quark expansion is 
factorially divergent. 

For the self-energy in the $\sigma$-model it is straightforward to 
derive from the formulae collected in the appendix that the contribution of 
large $n\gg t$ (for fixed $t$) to the 
sum in (\ref{SE}) is equal to 
\begin{equation}
\frac{9^{5/4}\,m^4}{\pi\,p^2}\,
\sum_{n\gg t} \,\frac{1}{n^2}\left(-\frac{9m^2}{p^2}\right)^n
\Bigg\{ \left(\frac{9m^2}{p^2}\right)^t
\Big[-\ln(9m^2/p^2)+\psi(t)-\psi(1-t)\Big]
-\Gamma(t)\Gamma(1-t)\Bigg\}.
\end{equation} 
It follows that the sum converges pointwise in $t$ for $|p^2|>9m^2$ and 
uniformly on any interval $[0,t_0]$. The contribution to 
$\Sigma(p)$ from $t>t_0$ can be made arbitrarily small by increasing 
$t_0$ as can be seen from inserting (\ref{at}) with lower integration 
limit $t_0$ into (\ref{sig}). Hence the domain of convergence of the 
OPE (i.e. the series after $t$-integration) is given by $|p^2|>9m^2$.


The radius of convergence seems to be related to the fact that the 
self-energy graph has a discontinuity for $-p^2>9 m^2$, related to the 
fact that any cut contains three $\sigma$-particles of mass $m$. We 
suggest that the convergence of the OPE is a consequence of the 
$1/N$ expansion. At every finite order of the $1/N$ expansion, a
cut can contain only a finite number of particles. We expect 
that the OPE of the self-energy is convergent 
for $|p^2|>(2k+1)^2 m^2$ if the $1/N$ expansion is truncated 
at the order $k$. For finite $N$, when there is no restriction on the 
number of particles in a cut, we expect that the OPE diverges for any 
value of $p^2$ in agreement with the conclusion of \cite{Shi94}. 
It may be interesting to pursue further the 
connection between the divergence of the OPE and multi-particle 
production. 

Another consequence of the absense of multi-particle thresholds is that, 
viewed as a function of $g$, the self-energy 
has no other singularities in $g$ 
than cuts stretching from $g=0$ to $g=1/(\ln 9+i n \pi)$ ($n$ a non-zero 
integer) 
in the complex coupling plane. This differs from QCD and the 
$\sigma$-model at finite $N$, where the analyticity domain has zero opening
angle \cite{thooft}, because in the presence of 
multi-particle thresholds up to arbitrarily high energies the 
singular points accumulate at $g=0$ on 
curves with zero opening angle at $g=0$.\\ 

\noindent
{\bf 5.}
In conclusion, the expansion of the self-energy in the $\sigma$-model 
provides the first analytic example of an OPE to all orders in $m^2/p^2$ 
in the non-trivial situation that each term in the expansion is multiplied 
by an  infinite series in the coupling that contains UV and IR 
renormalons. We discussed the cancellation of IR renormalon ambiguities 
in this expansion. We found that the OPE is convergent in the 
$1/N$-expansion and converges to the exact result without the need 
for additional power corrections. 

In \cite{latt} an unexpected $\Lambda^2/p^2$-correction (where 
$\Lambda$ is the QCD scale) is reported in the OPE of the plaquette 
expectation value. The present analysis does not offer an explanation 
of this fact. One difference is that the lattice calculation is done 
at finite UV cut-off $\Lambda_{UV}=a^{-1}$ with $p$ and $\Lambda_{UV}$ 
being identified. One then obtains additional power corrections from 
higher-dimension operators to the action. We can mimic this effect 
by adding the dimension-4 operator $C/\Lambda_{UV}^2\cdot
\sigma^2\partial^2\alpha$ to the $\sigma$-model action. The new vertex 
contributes an additional factor $k^2/\Lambda_{UV}^2$ to the integral 
(\ref{sig}). For $\Lambda_{UV}\sim p$ this results in an order 
1 contribution to regions (i,iii) and an order $m^6/p^6$ contribution  
to region (ii). The analogous argument for Yang-Mills theory shows 
that one cannot obtain a $\Lambda^2/p^2$-contribution in this way. The 
only loophole is that for $\Lambda_{UV}\sim p$ one should consider the 
infinite series of higher-dimension operators. If this series is 
divergent, this may induce further  
power correction terms in the lattice theory 
at finite cut-off not present in the continuum theory.

\vspace*{0.5cm}

\noindent
{\bf Acknowledgements.}
We thank G.~Grunberg and G.~Marchesini for useful discussions. 
The work by  N.~K. was partially supported by NORDITA through the 
Baltic Fellowship program funded by the Nordic Council of 
Ministers.\\


\noindent
{\bf Appendix.} We collect explicit expressions for the coefficients 
in the asymptotic expansion of the self-energy in (\ref{SE}). For $n=0$:
\begin{eqnarray}
F_{\rm p}^{(0)}[t]&=&1,
\nonumber\\
G_{\rm p}^{(0)}[t]&=& \frac{1}{t}+\frac{1}{t-1}
      -\psi(1+t)-\psi(2-t)-2\gamma_E,
\nonumber\\
H_{\rm np}^{(0)}[t]&=& \frac{1}{t}+B_1(t).
\end{eqnarray}
For $n=1$:
\begin{eqnarray}
F_{\rm p}^{(1)}[t]&=&t^2-1,
\nonumber\\
G_{\rm p}^{(1)}[t]&=&
-\frac{1}{t}+2t^2-4t+1+
   (1-t^2) \Big[\psi(1+t)+\psi(2-t)+2\gamma_E\Big],
\nonumber\\
H_{\rm np}^{(1)}[t]&=&
-\left(\frac{1}{t}+\frac{1}{t-1}+\frac{1}{t+1}-B_0(t)\right),
\end{eqnarray}
where $\psi(x)$ is the digamma-function and
$B_0$ and $B_1$ are scheme-dependent subtraction constants which are analytic 
in $t$: 
\begin{eqnarray}
   B_0(t) &=& 3 J(t,1),\qquad
   B_1(t) = 6J(t,2)-6J(t,3)+6J(1+t,3)-7J(1+t,2),
\nonumber\\
   J(t,m) &=& \frac{1}{t+1}\,{}_3F_2\left[
\left.
\begin{array}{l} 
1,m,t+1 \\
1+\frac{t}{2}, \frac32+\frac{t}{2}
\end{array}
\right|\frac14
\right]
\end{eqnarray}
For generic $n\ge 2$ we obtain:
\begin{eqnarray}
F_{\rm p}^{(n)}[t]&=& \sum_{k=0}^n K_{\rm p}(n-k,t)
  \frac{(n-k-1+t)^2_k}{k!k!},
\nonumber\\
H_{\rm np}^{(n)}[t]&=& \sum_{k=0}^{n-2} K_{\rm np}(n-k-2,t)
  \frac{(n-k-1)^2_k}{k!k!},
\\
G_{\rm p}^{(n)}[t]&=& \sum_{k=0}^n K_{\rm p}(n\!-\!k,t)
  \frac{(n\!-\!k\!-\!1\!+\!t)^2_k}{k!k!}\Big[
   \psi(t)-\psi(1-t)+2\psi(k+1)-2\psi(n+t-1)\Big]
\nonumber
\end{eqnarray}
where $(z)_k\equiv \Gamma(z+k)/\Gamma(z)$ and 
\begin{eqnarray}
 K_{\rm p}(n,t)&=&\frac1{n!}\frac{\Gamma(2t+2n-1)}{\Gamma(2t+n-1)}
\left[1-\frac{2n}{2t+n-1}+\frac{n(n-1)}{(2t+n)(2t+n-1)} \right],
\\
 K_{\rm np}(n,t)&=&\frac{\Gamma(-t,1+t,2n+3)}{\Gamma(t+n+1,3+n-t)}
\left[1-2\frac{n+2-t}{t+n+1}+\frac{(t-2-n)(t-1-n)}{(t+n+1)(t+n+2)} \right].
\nonumber
\label{K}
\end{eqnarray}
In the last formula we used a shorthand notation $\Gamma(a,b,c,\ldots)$
 for the product of $\Gamma$-functions with the respective arguments.

\end{document}